\documentclass[12pt]{article}
\usepackage[cp1251]{inputenc}
\usepackage{amsfonts}
\usepackage{latexsym}
\begin{document}

\begin{center}
{\Large \bf Constructing stabilized brane world models in
five-dimensional Brans-Dicke theory} \\

\vspace{4mm}

Alexey S. Mikhailov$^a$, Yuri S. Mikhailov$^a$, Mikhail N.
Smolyakov$^b$, Igor P. Volobuev$^b$\\ \vspace{0.5cm} $^a$Faculty
of Physics, Moscow State University\\ 119992 Moscow, Russia\\
$^b$Skobeltsyn Institute of Nuclear Physics, Moscow State
University
\\ 119992 Moscow, Russia \\
\end{center}
\begin{abstract}
We consider brane world models, which can be constructed in the
five-dimensional Brans-Dicke theory with bulk scalar field potentials
suggested by the supergravity theory. For different choices of the
potentials and parameters we get: (i) an unstabilized model with
the Randall-Sundrum solution for the metric and constant solution
for the scalar field; (ii) models with flat background and
tension-full branes; (iii) stabilized brane world models, one of
which reproduces the Randall-Sundrum solution for the metric and
gives an exponential solution for the scalar field. We also
discuss the relationship between solutions in different frames -- with
non-minimal and minimal coupling of the scalar field.
\end{abstract}
\section{Introduction}
Brane world models and their phenomenology have been widely
discussed in the last years \cite{1,11,12,13,14,Rubakov:2001kp}.
One of the most interesting brane world models is the
Randall-Sundrum model with two branes, -- the RS1 model
\cite{Randall:1999ee}. This model solves the hierarchy problem due
to the warp factor in the metric and predicts an interesting  new
physics in the TeV range of energies. A flaw of the RS1 model  is
the presence of a massless scalar mode, -- the radion, which
describes fluctuations of the branes with respect to each other.
As a consequence, one gets   a scalar-tensor theory of gravity on
the branes, the scalar component being described by the radion.
The radion coupling to matter on the negative tension brane
contradicts the existing restrictions  on the scalar component of
the gravitational interaction (see, for example, review
\cite{Bertolami:2006js}), and in order the model be
phenomenologically acceptable the radion must acquire a mass. The
latter is equivalent to the stabilization of the brane separation
distance, i.e. it must be defined by the model parameters. The
models, where the interbrane distance is fixed in this way, are
called stabilized models, unlike the unstabilized models, where
the interbrane distance can be arbitrary.

Such a  stabilization of the interbrane  distance can be achieved, for
example, by introducing a five-dimensional scalar field with brane
potentials \cite{wise}. A disadvantage of this approach  is that
the backreaction of the scalar field on the background metric is
not taken into account. This problem is solved in the well-known
model proposed in \cite{wolfe}. However the background metric of
this solution is rather complicated and differs significantly from
the Randall-Sundrum solution. An interesting problem is, whether
it is possible to find a stabilized model, where  the
Randall-Sundrum form of the metric is retained despite the
interaction with the scalar field.

A solution to this problem  based on  a non-minimal coupling of
the scalar field to gravity was put forward in \cite{Grzadkowski},
where the form of  the metric was found to be  the Randall-Sundrum
one, whereas the solution for the scalar field again turned out to
be rather complicated.

One of the standard forms of the non-minimal coupling of the
scalar field to gravity is the linear coupling to scalar curvature
used in the Brans-Dicke theory of gravity. Brans-Dicke theory in
the brane world context was already discussed in the literature,
see, for example, \cite{BD,BD1,BD2}.

In the present paper we use this type of coupling to construct a
stabilized model with the Randall-Sundrum solution for the metric,
the solution for the scalar field being also a simple exponential
function. Our paper is organized as follows. First, we present a
method for constructing different background solutions in
five-dimensional Brans-Dicke theory by considering bulk and brane
scalar field potentials of a special form and examine their
correspondence with the solutions in the Einstein frame. In
particular, we show that the solutions with these special scalar
field potentials in the Jordan frame correspond to the solutions
in the Einstein frame  with potentials
suggested by the supergravity theory. Then we discuss models which
can be obtained with different choices of the potentials and
parameters. And finally, we discuss the obtained results.

\section{Setup}
Let us denote the coordinates  in five-dimensional space-time
$E=M_4\times S^{1}/Z_{2}$  by $\{ x^M\} \equiv \{x^{\mu},y\}$, $M=
0,1,2,3,4, \, \mu=0,1,2,3 $, the coordinate $x^4 \equiv y$,
$-L\leq y \leq L$ parameterizing the fifth dimension. It forms the
orbifold, which is realized as the circle of the circumference
$2L$ with the points $y$ and $-y$ identified. Correspondingly, the
metric $g_{MN}$  and the scalar field $\phi$ satisfy the orbifold
symmetry conditions
\begin{eqnarray}
\label{orbifoldsym}
 g_{\mu \nu}(x,- y)=  g_{\mu \nu}(x,  y), \quad
  g_{\mu 4}(x,- y)= - g_{\mu 4}(x,  y), \\ \nonumber
   g_{44}(x,- y)=  g_{44}(x,  y), \quad
   \phi(x,- y)=  \phi(x,  y).
\end{eqnarray}
The branes are located at the fixed points of the orbifold, $y=0$
and $y=L$.

The  action of the brane world models can be written as
\begin{eqnarray}\label{actionDW}
S&=& \int d^{4}x \int_{-L}^L dy\sqrt{-g}\left[\phi R
-\frac{\omega}{\phi}g^{MN}\partial_M\phi\partial_N\phi-V(\phi)\right]
-\\ \nonumber &-&\int_{y=0} \sqrt{-\tilde g}\lambda_1(\phi) d^{4}x
-\int_{y=L}\sqrt{-\tilde g}\lambda_2(\phi)d^{4}x .
\end{eqnarray}
Here $V(\phi)$ is a bulk scalar field potential and
$\lambda_{1,2}(\phi)$ are brane scalar field potentials,
$\tilde{g}=det\tilde g_{\mu\nu}$, $\tilde g_{\mu\nu}$  denotes the
metric induced on the branes and $\omega$ is the five-dimensional
Brans-Dicke parameter. The formal restriction on this parameter in
the five-dimensional Brans-Dicke theory is $\omega>-\frac{4}{3}$
(which can be easily seen, for example, from the formulas of paper
\cite{conform}), and further restrictions on its value can be
obtained only after studying an effective four-dimensional theory
for a certain background solution. The only difference from the
classical Brans-Dicke theory is the presence of the bulk scalar
field potential $V(\phi)$ and the branes.  The signature of the
metric $g_{MN}$ is chosen to be $(-,+,+,+,+)$.

The standard ansatz  for the  metric and the scalar field, which
preserves the Poincar\'e invariance in any four-dimensional
subspace $y=const$, looks like
\begin{eqnarray}\label{metricDW}
ds^2&=&  e^{2\sigma(y)}\eta_{\mu\nu}  {dx^\mu  dx^\nu} +
dy^2,\quad \phi(x,y)=\phi(y),
\end{eqnarray}
 $\eta_{\mu\nu}$ denoting the flat Minkowski metric. If one
substitutes this ansatz into the equations corresponding to action
(\ref{actionDW}), one gets a rather complicated system of
nonlinear differential equations for functions
$\sigma(y),\phi(y)$:
\begin{eqnarray}\label{eq5}
& &3\sigma''\phi
+\frac{\omega}{\phi}(\phi')^{2}+\phi''-\sigma'\phi'+\frac{\lambda_1}{2}\delta(y)
+\frac{\lambda_2}{2}\delta(y-L)= 0,\\\label{eq6}&
&6(\sigma')^{2}\phi -
\frac{1}{2}\left(\frac{\omega}{\phi}(\phi')^{2}-V\right)+
4\sigma'\phi'=0,\\\label{eq7} & &
\frac{\omega}{\phi}\left(\phi''+4\sigma'\phi'\right)-4\sigma''-10(\sigma')^{2}
-\frac{\omega}{2\phi^{2}}(\phi')^2=\\ \nonumber
&=&\frac{1}{2}\frac{d V}{d\phi}+\frac{1}{2}\frac{d
\lambda_1}{d\phi}\delta(y)+\frac{1}{2}\frac{d
\lambda_2}{d\phi}\delta(y-L),
\end{eqnarray}
where prime denotes the derivative with respect to extra dimension
coordinate $y$.

We will consider a special class of bulk potentials,  namely
\begin{equation}\label{tildev}
V(\phi)=-(3\omega+4)\left[\left(4\omega+5\right)\phi
F^{2}+2\phi^{2} F\frac{d F}{d\phi}-3\phi^{3}\left(\frac{d
F}{d\phi}\right)^2\right],
\end{equation}
where $F\equiv F(\phi)$ is a function. This structure of the
potentials is similar to the one introduced in \cite{Brandhuber}
for the case of the minimal coupling of the scalar field to
gravity. Although the potential in its general form (\ref{tildev})
looks rather complicated, the resulting potentials for different
choices of the function $F$, as we will see below, look quite
natural.

One can check that in this case any solution of the equations
\begin{equation}\label{varphi}
\phi'=\phi F-3\phi^{2}\frac{d F}{d\phi},
\end{equation}
\begin{equation}\label{sigma}
\sigma'=\left(\omega+1\right)F+\phi\frac{d F}{d\phi}
\end{equation}
also satisfies (\ref{eq5}), (\ref{eq6}) and (\ref{eq7}) in the
interval $\left[0,L\right]$, provided that the following boundary
conditions on the branes are satisfied
\begin{equation}\label{bound1}
\left(3\omega+4\right)F|_{y=+0}=-\frac{\lambda_{1}}{4\phi},
\end{equation}
\begin{equation}\label{bound2}
\left(3\omega+4\right)F|_{y=L-0}=\frac{\lambda_{2}}{4\phi},
\end{equation}
\begin{equation}
\left(3\omega+4\right)\frac{d
F}{d\phi}|_{y=+0}=-\frac{1}{4}\frac{d\left(\lambda_{1}/\phi\right)}{d\phi},
\end{equation}
\begin{equation}\label{bound4}
\left(3\omega+4\right)\frac{d
F}{d\phi}|_{y=L-0}=\frac{1}{4}\frac{d\left(\lambda_{2}/\phi\right)}{d\phi}.
\end{equation}
It is necessary to note that the symmetry conditions
(\ref{orbifoldsym}) were used to obtain these relations.

Thus we get first order differential equations instead of the
initial second order differential equations. The situation is
analogous to that in papers \cite{wolfe,Brandhuber}, where the
bulk and the brane potentials for the scalar field minimally
coupled to five-dimensional gravity are chosen in an appropriate
way.

It is a common knowledge that action (\ref{actionDW}), called the
action in the Jordan frame, can be brought by a conformal
transformation to an action with the scalar field minimally
coupled to gravity, which is called the action in the Einstein
frame. Indeed, if in action (\ref{actionDW}) we put $\phi=2M^3
\exp(-\frac{3\rho}{a})$ with $a=6 M^{3/2}(\omega + 4/3)^{1/2}$ and
make a conformal rescaling $g_{MN}\rightarrow
\exp(\frac{2\rho}{a})g_{MN}$ allowing one to pass from the Jordan
frame to the Einstein frame, we get a model, describing scalar
field $\rho$ minimally coupled to five-dimensional gravity, with
the action
\begin{eqnarray}\label{actionsDW}
S&=& \int d^{4}x \int_{-L}^L dy \sqrt{-g}\left(2 M^3 R -
\frac{1}{2} g^{M N}\partial_M\rho\partial_N\rho - \bar
V(\rho)\right) - \\ \nonumber & - &\int_{y=0} \sqrt{-\tilde
g}\bar\lambda_1(\rho)  d^{4}x - \int_{y=L}\sqrt{-\tilde
g}\bar\lambda_2(\rho)    d^{4}x ,
\end{eqnarray}
where
$$
\bar V(\rho) = e^{\frac{5\rho}{a}}V(2M^3 e^{-\frac{3\rho}{a}}),
\quad \bar\lambda_i(\rho)= e^{\frac{4\rho}{a}}\lambda_i(2M^3
e^{-\frac{3\rho}{a}}), \quad i=1,2.
$$
Thus, if we a have a solution in the Jordan frame
$$
ds^2= e^{2\sigma(y)}\eta_{\mu\nu} dx^\mu dx^\nu + dy^2, \quad
\phi(x,y)= \phi(y),
$$
generated by potential (\ref{tildev}), it can be transformed to a
solution of the corresponding theory (\ref{actionsDW}) in the
Einstein frame
$$
ds^2= e^{2\left(\sigma(y)- \frac{\rho(y)}{a}\right)} \eta_{\mu\nu}
dx^\mu dx^\nu + e^{-\frac{2\rho}{a}}dy^2, \quad \rho(y)=
-\frac{a}{3} \ln\left(\frac{\phi(y)}{2M^3}\right).
$$
To bring the metric of this solution to the standard form
(\ref{metricDW}), we have to pass to a new coordinate $z$ of the
extra dimension according to $dz = \exp(-\frac{\rho(y)}{a}) dy$.
Thus, we get a solution in the Einstein frame
$$
ds^2= e^{2\left(\sigma(z)- \frac{\rho(z)}{a}\right)} \eta_{\mu\nu}
dx^\mu dx^\nu + dz^2, \quad \rho(z)=\rho(y(z)),
$$
in theory (\ref{actionsDW}) with potentials
\begin{eqnarray}\label{bulkpot-DW}
\bar V(\rho(z))&=& \frac{1}{8} \left(\frac{dW}{d\rho}\right)^2-\frac{1}{24M^{3}} W^2, \\ \label{F-W}
W(\rho(z))&=&-\frac{2a^{2}}{3}e^{\frac{\rho(z)}{a}}F(2M^3 e^{-\frac{3\rho}{a}}),\\
\label{branepot-DWF}
 \bar\lambda_i(\rho(z))&=& \pm W(\rho(z))|_{z=z_{i}}, \quad i=1,2,
\end{eqnarray}
which have exactly the same form as the potentials of the model
discussed in \cite{wolfe,Brandhuber}, where their form was
suggested by the supergravity theory. Thus, the method of finding
solutions in the Brans-Dicke theory is equivalent to the one
discussed in these papers, which is quite clear, because in both
cases the second order differential equations are reduced to the
first order ones. Nevertheless, if we study a Brans-Dicke theory,
it is more convenient to look for solutions in the Jordan frame
than to transform them from the Einstein frame.

 Another point is that solutions in one frame (Jordan or
Einstein) may be elegant and interesting in view of the hierarchy
problem, whereas in the other frame it is not the case. For
example, if in the Jordan frame we take polynomial functions $F$,
in the Einstein frame we get solutions for theories with
complicated potentials being a sum of several exponential
functions. Of course, the converse is also true. For this reason
it is also convenient to have a method of constructing brane world
solutions directly in the Jordan frame.

One more important point is that though solutions in different
frames are equivalent, the corresponding theories in different
frame are not equivalent, because  if in the Brans-Dicke theory we
had the minimal coupling of gravity to matter on the branes, then
after the conformal transformation the scalar field would enter
the term describing the interaction with matter on the branes (the
so-called conformal ambiguity \cite{Overduin:1998pn}). At the same
time, since the model described in \cite{wolfe} is stable against
fluctuations of gravitational and scalar fields, i.e. it is
tachyon- and ghost-free (see
\cite{BMSV}), the model described by (\ref{actionDW}) and
(\ref{tildev}) should possess this property too, although  the bulk
potentials for certain choices of $F$ are unbounded from below.

\section{Specific examples}
\subsection{The Randall-Sundrum solution}
Let us choose the function $F(\phi)$ in the following form
\begin{equation}
F=B\phi^{\frac{1}{3}},
\end{equation}
where $B$ is a constant. It follows from (\ref{varphi}) that
$\phi=const$ in this case. Bulk and brane potentials are
\begin{equation}\label{rsV}
V(\phi)=-\frac{4}{3}B^{2}\phi^{\frac{5}{3}}\left(3\omega+4\right)^{2},
\end{equation}
\begin{equation}\label{braneRSpotent}
\lambda_{1,2}=\mp 4B\left(3\omega+4\right)\phi^{\frac{4}{3}},
\end{equation}
and solution for the warp factor is
\begin{equation}
\sigma'=\frac{B}{3}\left(3\omega+4\right)\phi^{\frac{1}{3}}
\end{equation}
in the interval $[0,L]$. Making redefinition
\begin{equation}
\frac{\sqrt{2}}{3}B\left(3\omega+4\right)\phi^{\frac{1}{3}}=-k,
\end{equation}
\begin{equation}\label{phiRS}
\phi=2M^{3},
\end{equation}
we get
\begin{equation}\label{VphiRS}
V(\phi)=-12k^{2}M^{3},
\end{equation}
\begin{equation}
\lambda_{1,2}=\pm 12kM^{3},
\end{equation}
which formally coincide with the original RS1 solution
\cite{Randall:1999ee}. Nevertheless the linearized theory in this
background differs from that in the RS1 model, because the
fluctuations of the scalar field add an extra degree of freedom.
In this case the model cannot be stabilized (i.e. the size of the
extra dimension cannot be fixed) by adding positively defined
potentials on the branes, since the solution for the scalar field
does not depend on $y$ (this point is discussed in detail in
subsection~3.3). Thus, this example is not interesting from the
physical point of view, but it shows how the general method for
constructing solutions works.

It is not difficult to see that action (\ref{actionDW}) with
potentials of the form (\ref{rsV}) and (\ref{braneRSpotent}) can
be brought by a conformal transformation of the metric and
coordinate transformations, described in Section~2, to the action
of the unstabilized RS1 model with minimally coupled scalar field.
The only significant difference is that if in the Brans-Dicke
theory we had the minimal coupling of gravity to matter on the
branes, then after the transformation the scalar field would enter
the term describing the interaction with matter on the branes.

\subsection{"Consistent" ADD scenario}
Let us consider the case, where $\omega=0$, i.e. the kinetic term
for the Brans-Dicke scalar field is absent. Let us also suppose
that $F=B/\phi$. It is not difficult to check that
\begin{equation}
V(\phi)\equiv 0,
\end{equation}
\begin{equation}\label{sigmaADD}
\sigma'=0
\end{equation}
in this case. Equation (\ref{sigmaADD}) means that we have the
flat five-dimensional background metric. Solution for $\phi$,
following from (\ref{varphi}), has the form
\begin{equation}
\phi=4By+D
\end{equation}
in the interval $[0,L]$, where $D$ is a constant. From the
boundary conditions one gets
\begin{equation}
\lambda_{1,2}=\mp\lambda,
\end{equation}
where $\lambda$ is a constant defining the brane tensions. Finally
we get
\begin{equation}
\phi=\frac{\lambda}{4}|y|+D.
\end{equation}
Thus, we get the model with the flat five-dimensional background
metric and tension-full branes, which was discussed in detail in
paper \cite{Smol}. This model can be easily stabilized by the same
method as the one, which will be discussed in the next subsection.
\subsection{Stabilized brane world with the Randall-Sundrum solution for the metric}
Now let us consider the case $F=const$, i.e. it does not depend on
the field $\phi$. Bulk and brane potentials, corresponding to such
a choice of $F(\phi)$, can be chosen to be (see (\ref{tildev}) and
(\ref{bound1})--(\ref{bound4}))
\begin{equation}
V(\phi)=\Lambda\phi,
\end{equation}
\begin{equation}
\lambda_{1,2}=\pm\lambda\phi.
\end{equation}
Let us suppose that $\lambda>0$. It is not difficult to check that
from (\ref{varphi})--(\ref{bound4}) follows
\begin{equation}\label{sigmastabil}
\sigma=-k|y|,
\end{equation}
\begin{equation}\label{phistabil}
\phi=Ce^{-u|y|}
\end{equation}
with
\begin{equation}
u=\sqrt{\frac{-\Lambda}{\left(3\omega+4\right)\left(4\omega+5\right)}},
\end{equation}
\begin{equation}
k=(\omega+1)u,
\end{equation}
and the fine-tuning relation
\begin{equation}\label{ftune}
\lambda=4\sqrt{-\Lambda}\sqrt{\frac{3\omega+4}{4\omega+5}}.
\end{equation}
Constant $C$ is not defined by the equations. One can see that in
the limit $\omega\to\infty$ we arrive at the standard
Randall-Sundrum solution.

Now let us discuss stabilization mechanism which can be utilized
in the case under consideration. We will follow the way proposed
in \cite{wolfe} and add stabilizing quadratic potentials on the
branes, namely
\begin{equation}\label{higgs}
\Delta\lambda_{1,2}=\gamma_{1,2}\left(\phi-v_{1,2}\right)^{2},
\quad \gamma_{1,2}>0.
\end{equation}
Such an addition will not affect equations of motion provided
\begin{equation}
\phi|_{y=0}=v_{1},\quad \phi|_{y=L}=v_{2}.
\end{equation}
Thus, now the constant $C$ appears to be defined and  is equal
$C=v_{1}$, whereas the size of extra dimension is now defined by
the relation
\begin{equation}
L=\frac{1}{u}\ln\left(\frac{v_{1}}{v_{2}}\right),
\end{equation}
which is the same as in the model of paper \cite{wolfe}.
Obviously, this stabilization mechanism works only for solutions
with the scalar field depending on the extra coordinate, in
particular, for the solution in subsection~3.2, and does not work
for the solution in subsection~3.1. We would like to emphasize
that this mechanism  differs somewhat from the Goldberger-Wise
mechanism \cite{wise}, because it takes into account the
backreaction of the scalar field on the metric, and the size of
the extra dimension is fixed by the  boundary values of the
former.

Now let us find the relationship between the four-dimensional
Planck mass and the parameters of the theory. We assume that the
brane at $y=L$ is "our"\ brane. To this end one should choose
$\sigma=-k|y|+kL$ to make four-dimensional coordinates
$\{x^{\mu}\}$ Galilean on this brane (this problem was discussed
in detail in \cite{BKSV}). Naive considerations suggest (see, for
example, \cite{Smol}) that the wave function of the massless
four-dimensional tensor graviton has the same form as that in the
unstabilized Randall-Sundrum model, namely
$h_{\mu\nu}^{0}(x,y)=e^{2\sigma}\alpha_{\mu\nu}(x)$. Moreover, the
form of the residual gauge transformation, which are left after
imposing the gauge on the fluctuations of metric (see
\cite{BMSV,BKSV}), also suggests the same form of the wave
function. Thus, substituting the following ansatz
\begin{eqnarray}
g_{\mu\nu}(x,y)=e^{2\sigma}g^{(4)}_{\mu\nu}(x), \quad
\phi(x,y)=\phi(y),\\ \nonumber g_{44}(x,y)=1, \quad g_{\mu
4}(x,y)=0
\end{eqnarray}
into action (\ref{actionDW}), we get
\begin{equation}\label{effPlanck}
S=\int_{-L}^{L}\phi e^{2\sigma}dy \int
R_{(4)}\sqrt{-g_{(4)}}\,d^{4}x
\end{equation}
and
\begin{equation}\label{Planck1}
2M^{2}_{Pl}=\int_{-L}^{L}\phi e^{2\sigma}
dy=\frac{2v_{1}}{2k+u}\left(e^{2kL}-e^{-uL}\right).
\end{equation}
If $uL<1$ we get a formula analogous to that in the unstabilized
Randall-Sundrum model (see \cite{Rubakov:2001kp,BKSV}). To solve
the hierarchy problem one needs $kL$ to be of the order $kL\sim
30$. If one chooses relatively large $\omega$ (for example,
$\omega\ge 30$), then $k$ would go to the value corresponding to
the unstabilized Randall-Sundrum model, namely
\begin{equation}
k\approx \sqrt{\frac{-\Lambda}{12}},
\end{equation}
(compare with (\ref{phiRS}) and (\ref{VphiRS})) whereas $uL$ could
be made less than unity ($uL<1$), since $u=k/(\omega+1)$. Under
this assumption the parameter $v_{1}$ can be chosen to be of the
same order as $v_{2}$. In this case the parameters of the model,
made dimensionless with the help of a fundamental scale in the
$TeV$ range, do not contain a hierarchical difference. The
situation turns out to be completely analogous to that in the
model proposed in \cite{wolfe}. At the same time, solution for the
warp factor in stabilized brane world model found in \cite{wolfe}
is quite complicated, which impedes the analysis of the equations
of motion for linearized gravity (approximate solutions can be
found  only under certain assumptions and simplifications, see
\cite{BMSV}). As for our case, one can think that though the
general structure of  action (\ref{actionDW}) is more complicated
than that of the action used in \cite{wolfe}, the simplicity of
solutions (\ref{sigmastabil}) and (\ref{phistabil}) could result
in simpler equations of motion for linearized gravity.

Another advantage of the solution presented above is that one can
use all the results, obtained for the case of the universal extra
dimensions in the Randall-Sundrum model (i.e. if one allows
additional fields to propagate in the bulk, see, for example,
\cite{Burdman} and references therein), in our case too. This
happens because of the equivalence of solutions for the warp
factors in both models. Of course, it is true in the case of the
standard coupling of five-dimensional gravity to matter in the
bulk. But since the size of extra dimension in our model appears
to be stabilized,  one can think that this would allow us to avoid
possible problems caused by the radion field, which are inherent
in the unstabilized Randall-Sundrum model.

Quite an interesting situation arises, if one chooses $\omega=-1$.
Although the Brans-Dicke parameter is negative, the model is
stable, since $\omega>-\frac{4}{3}$. Formulas
(\ref{sigmastabil})--(\ref{ftune}) with $\omega=-1$ take the form
\begin{equation}
\sigma=0,\quad (k=0),
\end{equation}
which means that the five-dimensional background metric is flat,
\begin{equation}
\phi=Ce^{-u|y|},
\end{equation}
\begin{equation}
u=\sqrt{-\Lambda},
\end{equation}
\begin{equation}
\lambda=4\sqrt{-\Lambda}.
\end{equation}
Thus we get a model, which is similar, to some extent, to that
discussed in subsection~3.2. In order to have the hierarchy
problem solved (in the way proposed in \cite{Smol}) in case of
$TeV$ range of fundamental five-dimensional physics, one should
choose $uL\sim 30$, as in the Randall-Sundrum model, and $C\sim
e^{uL}$. The flaw of the case $\omega=-1$ is that there appears a
new hierarchy between $v_{1}$ and $v_{2}$. Nevertheless, this
choice of parameters can be interesting from the pedagogical point
of view since it demonstrates another scenario with flat
background and tension-full branes (and nonempty bulk).

The solutions of this section can be easily related to solutions
in the Einstein frame. For these solutions $F=const$, and $\bar V(\rho(z))$
(see (\ref{bulkpot-DW}) and (\ref{F-W})) is the Liouville
potential. Such "dilatonic"\ brane worlds were widely discussed in
the literature, see, for example \cite{Liouv,Liouv1}. As we have
shown, our solutions can be brought to the form of the general
solutions of paper \cite{Liouv} by a conformal rescaling of the
metric and an appropriate transformation of the extra dimension
coordinate, the latter being necessary for retaining the same
ansatz for the metric (of the form (\ref{metricDW})).

Our results also give a simple solution to the problem of
finding a stabilized brane world model with the Randall-Sundrum
form of the metric, which was discussed in \cite{Grzadkowski}. In
this paper a similar solution in a theory with the scalar field
non-minimally coupled to gravity, which preserves the
Randall-Sundrum form of the metric, was found. But this solution
and ours cannot be transformed to each other by a redefinition of
fields, rescaling of metric and coordinate transformations.

We would like to note once again that the theories obtained by a
rescaling of the metric are not equivalent, if one considers the
standard coupling of gravity to matter on the branes, as we have
already mentioned in the end of Section~2. Since in stabilized
brane world models the fluctuations of the scalar field describe
also the radion (see \cite{BMSV}), this ambiguity modifies, in
particular, the radion coupling to matter on the branes. Because
we do not know, which frame is the "real"\ one, there are no
strong objections against choosing the Jordan one. In this
connection, an interesting problem is to compare the physical
consequences of the models in different frames, which can be
transformed one into another in the absence of matter on the
branes.

\subsection{Power law solutions} Let us consider the
case
\begin{equation}
F=B\phi^{n},
\end{equation}
where $n\ne\frac{1}{3}$ and $n\ne 0$ (these cases were discussed
above). The corresponding bulk potential has the form
\begin{equation}
V(\phi)=-(3\omega+4)B^{2}\phi^{2n+1}\left[4\omega+5+2n-3n^{2}\right],
\end{equation}
whereas the fine-tuned brane potentials can be easily obtained
from (\ref{bound1})--(\ref{bound4}). Solutions for the scalar
field and the warp factor, following from
(\ref{varphi})--(\ref{bound4}), are
\begin{equation}\label{sigma-general}
\sigma=\frac{\omega+1+n}{\left(3n-1\right)n}\ln\left[n\left(3n-1\right)B|y|+C\right]+C_{1},
\end{equation}
\begin{equation}\label{phi-general}
\phi=\left[n\left(3n-1\right)B|y|+C\right]^{-\frac{1}{n}},
\end{equation}
where constants $C$ and $C_{1}$ are not defined by the equations.
We will consider only such values of parameters $n$, $B$ and $C$
that the expression $\left[n\left(3n-1\right)B|y|+C\right]$ is
positive for any value of $y$.

Let us suppose that that "our"\ brane is that at the point $y=0$
(not that at the point $y=L$, as in the previous example). In this
case we should take
\begin{equation}\label{c-one}
C_{1}=-\frac{\omega+1+n}{\left(3n-1\right)n}\ln\left(C\right)
\end{equation}
to make four-dimensional coordinates $\{x^{\mu}\}$ be Galilean on
this brane (i.e. $\sigma(y=0)=0$). The constant $C$ is defined by
the stabilizing potential on the brane at $y=0$, whereas the size
of the extra dimension is defined by the stabilizing potential on
the brane at $y=L$.

After some algebra we can easily get from (\ref{effPlanck}) the
value of effective four-dimensional Planck mass on the brane at
$y=0$
\begin{eqnarray}\label{effPlanck1-general}
M^{2}_{Pl}&=&\int_{0}^{L}\phi e^{2\sigma} dy=\\
\nonumber
&=&\frac{C^{\frac{n-1}{n}}}{B\left(2\omega+3+3n^{2}-2n\right)}\left[\left(\frac{n\left(3n-1\right)BL}{C}+1\right)^{\frac{2\omega+3+3n^{2}-2n}
{n\left(3n-1\right)}}-1\right].
\end{eqnarray}

We will show why such solutions can be interesting from the point
of view of hierarchy problem by utilizing the choice\footnote{The
authors are grateful to K.~Farakos and P.~Pasipoularides for
suggesting to examine this case and the corresponding background
solution, which resulted in this subsection}
\begin{equation}
n=\frac{3}{2}.
\end{equation}
Bulk and brane potentials, corresponding to such a choice of $n$,
have the form
\begin{equation}
V(\phi)=-(3\omega+4)\frac{16\omega+5}{4}B^{2}\phi^{4},
\end{equation}
\begin{equation}
\lambda_{1,2}=\mp 4B\left(3\omega+4\right)\phi^{\frac{5}{2}}.
\end{equation}
Solutions for the scalar field and the warp factor, following from
(\ref{sigma-general}), (\ref{phi-general}) and (\ref{c-one}), are
\begin{equation}
\sigma=\frac{4\omega+10}{21}\left[\ln{\left(21B|y|+D\right)}-\ln{\left(D\right)}\right],
\end{equation}
\begin{equation}
\phi=\left(\frac{4}{21B|y|+D}\right)
^{\frac{2}{3}},
\end{equation}
where $D$ is a constant which will be defined below.

To stabilize the size of the extra dimension, we add potentials of
the form (\ref{higgs}) on the branes:
\begin{equation}
\Delta\lambda_{1,2}=\gamma_{1,2}\left(\phi-v_{1,2}\right)^{2}.
\end{equation}
As in the case discussed in the previous subsection, we get
\begin{equation}\label{boundaryhiggs}
\phi|_{y=0}=v_{1},\quad \phi|_{y=L}=v_{2}.
\end{equation}
The constant $D$ appears to be defined by
\begin{equation}
D=4\left(v_{1}\right)^{-3/2},
\end{equation}
and we get
\begin{equation}\label{phinew}
\phi=v_{1}\left(\frac{1}{\frac{21\left(v_{1}\right)^{3/2}B|y|}{4}+1}\right)^{\frac{2}{3}}.
\end{equation}
It follows from (\ref{boundaryhiggs}) and (\ref{phinew}) that the
size of extra dimension is defined by the relation
\begin{equation}
L=\frac{4}{21B}\left[\frac{\left(v_{1}\right)^{3/2}-\left(v_{2}\right)^{3/2}}{\left(v_{1}v_{2}\right)^{3/2}}\right].
\end{equation}
The warp factor has the form
\begin{equation}\label{sigmanew}
e^{\sigma}=\left(\frac{21\left(v_{1}\right)^{3/2}B|y|}{4}+1\right)^{\frac{4\omega+10}{21}},
\end{equation}
and it is not difficult to calculate the value of the
four-dimensional Planck mass on the brane at $y=0$. Using
(\ref{effPlanck1-general}) we can easily get
\begin{equation}\label{effPlanck1}
M^{2}_{Pl}=\int_{0}^{L}\phi e^{2\sigma}
dy=\frac{4}{B\left(v_{1}\right)^{1/2}\left(8\omega+27\right)}\left[\left(\frac{v_{1}}{v_{2}}\right)^{\frac{8\omega+27}{14}}-1\right].
\end{equation}

Let us suppose that all the parameters of the model, made
dimensionless with the help of a fundamental scale in the $TeV$
range, do not contain a hierarchical difference. For example, one
can choose $B\approx 1 TeV^{-7/2}$, $v_{1}\approx 1 TeV^{3}$ and
$\frac{v_{1}}{v_{2}}\approx 3.4$. It follows from this assumption
that $B\left(v_{1}\right)^{3/2}L\approx 1$, which means that
$L\approx 1 TeV^{-1}$. If one chooses $\omega=110$, then the
four-dimensional Planck mass on the brane at $y=0$ appears to be
of the order of $M_{Pl}\sim 10^{19} GeV$. Thus, although the
largest dimensionless parameter in the model is $\omega=110$, we
get a difference in $16$ orders of magnitude between four- and
five-dimensional energy scales. We also see, that in this case the
hierarchy problem is solved because of the large power in
(\ref{effPlanck1}), contrary to the case discussed in the previous
subsection, where the hierarchy problem is solved due to the
exponential factor (as in the Randall-Sundrum model). Of course,
if we take larger value of $\frac{v_{1}}{v_{2}}$, the value of
$\omega$ can be much smaller.

\section{Conclusion and final remarks}
In this paper we considered five-dimensional Brans-Dicke theory as
a basis for building different solutions corresponding to brane
world models. It appeared that for  particular choices of the
potentials and certain values of parameters the theory reproduces
some known background solutions. We also presented new solutions
for stabilized brane worlds, one of which has a relationship with
a known solution in another frame. We hope that appropriate choice
of the function $F(\phi)$ can lead to other interesting solutions,
which are not evident at the first glance.

A reasonable question arises -- what happens to the mass of the
radion and its coupling to matter on the brane, especially in the
stabilized cases discussed in subsections~3.3, 3.4? It is clear
that the radion mass should be expressed through the model
parameters $\gamma_{1,2}$, $\Lambda$ (or $B$, respectively),
$\omega$ and $v_{1,2}$. Calculations made in \cite{BMSV} for the
stabilized brane world model proposed in \cite{wolfe} suggest that
with an appropriate choice of these parameters the radion mass can
be made to be in the $TeV$ range, which can be interesting from
the experimental point of view and does not contradict the known
data. Nevertheless, an answer to the question posed above can be
obtained only after a thorough examination of linearized gravity
in the models. This issue calls for further investigation.

We would like to note that all the models found in the present
paper have a finite size of extra dimension and are of interest
for solving the hierarchy problem. At the same time, it would be
interesting to look for "fat brane"\ solutions in five-dimensional
Brans-Dicke theory, analogous to that found in
\cite{Kehagias:2000au}, and to examine their properties. This
problem also deserves to be studied.

Finally, it is necessary to mention that there are other types of
non-minimal coupling of the scalar field to gravity. For example,
in  recent papers \cite{ricci0,ricci1,ricci2} some interesting
brane world solutions (both analytic and numerical) in the theory
with Ricci-coupled bulk scalar field were found.  In this
connection it would be interesting to compare the physical
consequences of the models with different types of non-minimal
coupling and identical solutions for the warp factor -- for
example, of the model discussed in subsection~3.3 of this paper
and of the models discussed in \cite{Grzadkowski,ricci0,ricci2},
in the case of the standard coupling of gravity to matter on the
branes in all the models.

\section*{Acknowledgments}
The authors would like to thank E.E.~Boos and Yu.V.~Grats for
valuable discussions. The authors are also grateful to K.~Farakos
and P.~Pasipoularides for many useful remarks. The work of M. S.
and I. V. was supported by RFBR grant 04-02-17448 and Russian
Ministry of Education and Science grant NS-8122.2006.2. M. S. also
acknowledges support of grant for young scientists MK-8718.2006.2
of the President of Russian Federation.


\begin{thebibliography}{99}
\bibitem{1}
F. Feruglio, {\em Eur. Phys. J. C} {\bf 33} (2004) S114

\bibitem{11}
S. Mele, {\em Eur. Phys. J. C} {\bf 33} (2004) S919

\bibitem{12}
I. Antoniadis, {\em Eur. Phys. J. C} {\bf 33} (2004) S914

\bibitem{13}
C. Csaki, "TASI lectures on extra dimensions and branes",
hep-ph/0404096

\bibitem{14}
J. Hewett, J. March-Russell, {\em Phys. Lett. B} {\bf 592} (2004)
1

\bibitem{Rubakov:2001kp}
V.A. Rubakov, {\em Phys. Usp.} {\bf 44} (2001) 871

\bibitem{Randall:1999ee}
L. Randall, R. Sundrum, {\em Phys. Rev. Lett.} {\bf 83} (1999)
3370

\bibitem{Bertolami:2006js}
O. Bertolami, J. Paramos, S.G. Turyshev, "General theory of
relativity: Will it survive the next decade?", gr-qc/0602016

\bibitem{wise}
W.D. Goldberger, M.B. Wise, {\em Phys. Rev. Lett.} {\bf 83} (1999)
4922

\bibitem{wolfe}
O. DeWolfe, D.Z. Freedman, S.S. Gubser, A. Karch, {\em Phys. Rev.
D} {\bf 62} (2000) 046008

\bibitem{Grzadkowski}
B. Grzadkowski, J.F. Gunion, {\em Phys. Rev. D} {\bf 68} (2003)
055002

\bibitem{BD}
L.E. Mendes, A. Mazumdar, {\em Phys. Lett. B} {\bf 501} (2001) 249

\bibitem{BD1}
L. Perivolaropoulos, {\em Phys. Rev. D} {\bf 67} (2003) 123516

\bibitem{BD2}
M. Arik, D. Ciftci, {\em Gen. Rel. Grav.} {\bf 37} (2005) 2211

\bibitem{conform}
V. Faraoni, E. Gunzig, P. Nardone, {\it Fund. Cosmic Phys.} {\bf
20} (1999) 121

\bibitem{Brandhuber}
A. Brandhuber, K. Sfetsos, {\it J. High Energy Phys.} {\bf 10}
(1999) 013

\bibitem{Overduin:1998pn}
J.M. Overduin, P.S. Wesson, {\it Phys. Rept.} {\bf 283} (1997) 303

\bibitem{BMSV}
E.E. Boos, Yu.S. Mikhailov, M.N. Smolyakov, I.P. Volobuev, {\it
Mod. Phys. Lett. A} {\bf 21} (2006) 1431

\bibitem{Smol}
M.N. Smolyakov, "Consistent ADD scenario with stabilized extra
dimension", hep-th/0507216

\bibitem{BKSV}
E.E. Boos, Yu.A. Kubyshin, M.N. Smolyakov, I.P. Volobuev, {\em
Class. Quant. Grav.} {\bf 19} (2002) 4591

\bibitem{Burdman}
G. Burdman, {\it AIP Conf. Proc.} {\bf 753} (2005) 390

\bibitem{Liouv}
S. Kachru, M.B. Schulz, E. Silverstein, {\em Phys. Rev. D} {\bf
62} (2000) 045021

\bibitem{Liouv1}
N. Alonso-Alberca, B. Janssen, P.J. Silva, {\em Class. Quant.
Grav.} {\bf 17} (2000) L163

\bibitem{Kehagias:2000au}
A. Kehagias, K. Tamvakis, {\em Phys. Lett. B} {\bf 504} (2001) 38

\bibitem{ricci0}
K. Farakos, P. Pasipoularides, {\em Phys. Lett. B} {\bf 621}
(2005) 224

\bibitem{ricci1}
K. Farakos, P. Pasipoularides, {\em Phys. Rev. D} {\bf 73} (2006)
084012

\bibitem{ricci2}
C. Bogdanos, A. Dimitriadis, K. Tamvakis, {\em Phys. Rev. D} {\bf
74} (2006) 045003

\end{thebibliography}
\end{document}